# A D-vine copula-based coupling uncertainty analysis method for the stiffness predication of variable-stiffness composite

Qidi Li, Hu Wang[Ψ], Yang Zeng, Zhiwei Lv

State Key Laboratory of Advanced Design and Manufacturing for Vehicle Body, Hunan University, China

Joint Center for Intelligent New Energy Vehicle, China

**Abstract**: This study suggests a coupling uncertainty analysis method to investigate the stiffness characteristics of variable-stiffness (VS) composite. The D-vine copula function is used to address the coupling of random variables. To identify the copula relation between random variables, a novel one-step Bayesian copula model selection (OBCS) method is proposed to obtain a suitable copula function as well as the marginal CDF of random variables. The entire process is Monte Carlo simulation (MCS). However, due to the expensive computational cost of complete finite element analysis (FEA) in MCS, a fast solver, reanalysis method is introduced. To further improve the efficiency of entire procedure, a back propagation neural network (BPNN) model is also introduced based on the reanalysis method. Compared with the reanalysis method, BPNN shows a higher efficiency as well as sufficient accuracy. Finally, the fiber angle deviation of VS composite is investigated by the suggested strategy. Two numerical examples are presented to verify the feasibility of this method.

# 1. Introduction

Composite material nowadays is playing an increasing important role in the field of aerospace, shipping and automotive due to its attractive characteristics of high strength, stiffness and

---

[Ψ] Corresponding author, wanghu@hnu.edu.cn

light weight. Therefore, the investigation of composite material draws an extensive attention from all over the world in the past decades. Tobias[1] developed a multi-field finite element approach to model the fiber-reinforce composite (FRC). Using the 3-D printing technology, Justo *et al*. [2] printed the FRC and conduct experiment to examine its mechanical characteristics. Result showed that 3-D printing seems to be a promising approach to print long fiber composite. Wu [3] developed a multiscale mean-field homogenization approach to study the damage model of FRC. Compared to the FRC, the curved fiber distribution would lead to a variability of the stiffness and the curvilinear-line composite laminar is known as variable-stiffness (VS) composite. Due to the variability of fiber angle orientation, the design of freedom is much higher and simultaneously, the difficulty is larger either. Shahriar [4] introduced a cellular automata based strategy to optimized the fiber placement in VS composite. Moreover, Wang and Huang [5] develop an optimized strategy using the reanalysis based finite element model as objective function and the genetic algorithm as optimization algorithm. Peng *et al* [6,7] also dedicated to the optimized design of the VS composite and moreover, they develop a isogeometric analysis (IGA) method to investigate the buckling of the VS composite.

Nowadays, the manufacturing of VS composite is based on an automatic fiber placing technology and this technology would inevitably lead to a fiber angle misalignment. Moreover, other defects of VS composite, such as the material properties uncertainty, matrix porosity, and overlaps, is also unavoidable during manufacturing. All these uncertainties would propagate to the variability of performance and it may even result in catastrophe in some places. Therefore, it is necessary to incorporated uncertainties while evaluating its full-scale performance.

In the past decade, numerous studies have been devoted to the uncertainty investigation of FRC. A perturbation based stochastic finite element has been introduced to uncertainty analysis focusing on the material properties [8, 9]. Innovatively, Cui and Hu [10] developed a copula-based perturbation method to study the uncertainty response of FRC and successfully derive the reliability index under different correlation coefficient. However, the drawback of the perturbation method, small perturbation range, limits its application in large perturbation cases. Moreover, Spectral stochastic finite element (SSFEM) was also included. Stefanou [11] and Chen [12] studied the static performance of the FRC by considering the uncertainty of elastic modulus. Recently, Sepahvand [13] used the SSFEM to investigate the nature frequency of the FRC by considering the fiber angle deviation. It obtained the probability distribution of the nature frequency as well as the vibration model successfully. In addition, Monte Carlo simulation (MCS) is another powerful and accurate uncertainty analysis method. It attracted a big attention in implementing static or dynamic analysis to FRC due to its accuracy and flexibility [14-16], but it is always time consuming and difficult to converge when degree of freedom (DOF) is large. Till now, most of the uncertainty analysis is focused on the FRC, but the uncertainty analysis of VS composite was not studied intensively. Moreover, most of the uncertainty analysis are focused on the case of independent random variables. Only a few works consider the coupling random variables. However, there are always some strong or weak correlation between random variables in practice. To this end, a vine copula-based coupling uncertainty analysis strategy is proposed trying to address this problem in this study.

Copula function is a powerful and flexible toolbox that connects the marginal probabilistic distribution functions of variables and then provides a joint distribution function of them [18]. It has a rich family to illustrate the relationship between two random variables, such as Clayton, AMH, Gumbel, Frank, Gauss, Joe and FGM Copula [19]. However, multivariate copula is difficult to derive directly. Fortunately, vine copula provides an effective way to address this problem. Vine copula constructs the multivariable joint distribution function by using pair-copula construction (PCC). Two majorities of vine copula are D-vine and C-vine copula and the D-vine copula is investigated in this study.

Though copula is an impressive tool in studying correlation between random variables, an inevitable problem is to determine the marginal probability distributions of random variables as well as their copula relation. Many kinds of methods devoted to solve this problem, such as Akaike information criteria (AIC), goodness of fit (GOF) and Bayesian method. Recently, model selection based on Bayesian method is increasingly popular due to its feasibility and flexibility. In this study, a novel one-step Bayesian copula selection method (OBCS) is introduced to handle this problem. Compared with the traditional Bayesian method, it identifies the marginal probability distribution as well as copula relation in a one-step process while the traditional Bayesian method has to finish this process in two steps. Therefore, OBCS can make full use of the input information as well as prohibit error propagation. Subsequently, based on the marginal distribution of random variables and their copula relation, a D-vine simulation is presented to derive a large sample for MCS.

MCS is one of the most widely used methods in uncertainty analysis due to its accuracy and flexibility. However, one prominent drawback of MCS is that the computational cost is considerably huge. Reanalysis method [21] can alleviate this problem in some extent. Different from the full analysis, reanalysis method focuses on the modified part of the system and only calculate the changing part. Therefore, the efficiency is highly improved compared to the full analysis [22-24]. Though reanalysis can improve the efficiency of MCS, the calculation process is still relatively time consuming. To improve the efficiency of entire program, a back propagation neural network (BPNN) model is introduced. The BPNN is a kind of feedforward neural network and widely used in the field of target identification and regression. It has a strong capability in training a nonlinear model [25,26].

In this study, fiber angle deviation of VS composite would be treated as the random variables and assumed to be correlated between plies. Moreover, a basic hypothesis is that the deviation of the fiber is assumed to be the same in a ply. Figure 1 shows the framework of this study. According to the experimental data, marginal probability distribution of random variables and copula correlation between them are identified through OBCS method and then samples are generated by the D-vine copula simulation. Subsequently, samples will be applied to the reanalysis assisted FEA and the system stiffness characteristic can be derived. Finally, based on the reanalysis method, the BPNN model can be trained. Using the well trained BPNN model, distribution of deformation of VS composite can be obtained. In addition, the efficiency of MCS can be extremely improved and the accuracy can be guaranteed.

The rest of this paper is organized as follow: in section 2, D-vine copula is introduced, including the OBCS method as well as the D-vine simulation method; in section 3, two evaluation methods of VS composite are introduced, which are reanalysis based FEA and BPNN surrogated method; Based on the suggested strategy, two numerical examples will be given in section 4 and the final conclusion will be included in section 5.

## 2. OBCS assisted D-vine simulation

### 2.1 D-vine copula

Copulas are *d*-dimensional multivariate distributions with uniformly distributed marginal distributions on [0, 1]. It is a powerful tool when modeling a dependence structure of multivariable.

Supposed that $X = (X_1, X_2, ..., X_d)^T$ be a *d*-dimensional random vector. It has joint distribution function of $F(x_1, x_2, ..., x_d)$ and marginal distribution $F_i(x_i), i = 1, 2, ..., d$. According to Sklar's theory [18], there exist a copula $C\{F_1(x_1), F_2(x_2), ..., F_d(x_d)\}$ such that

$$F(x_1, x_2, ..., x_d) = C\{F_1(x_1), F_2(x_2), ..., F_d(x_d)\} \tag{1}$$

If the marginal distribution $F_i(x_i)$ is continuous, then the multivariable distribution function determined by copula function is unique, which means that for different marginal distribution function, it might be a same copula to describe corresponding dependence structure.

Deviating both side of Eq. (1), the joint probability density function (PDF) can be expressed as a product of corresponding copula density and the marginal distribution function

$$f(x_1, x_2, ..., x_d) = c\{F_1(x_1), F_2(x_2), ..., F_d(x_d)\} \cdot f_1(x_1) f_2(x_2) ... f_d(x_d) \tag{2}$$

where $c\{F_1(x_1), F_2(x_2), ..., F_d(x_d)\}$ denotes the density of multivariate copula function and has the expression of

$$c\{F_1(x_1), F_2(x_2), ..., F_d(x_d)\} = \frac{\partial C\{F_1(x_1), F_2(x_2), ..., F_d(x_d)\}}{\partial F_1(x_1) \partial F_2(x_2) ... \partial F_d(x_d)} \tag{3}$$

It is difficult to derive $C\{F_1(x_1), F_2(x_2), ..., F_d(x_d)\}$ especially when the dimension is increased. Therefore, a method of pair-copula constructions (PCC) which firstly proposed by Aas *et al*. [27] is introduced. Let $f(x_1, x_2, ..., x_d)$ be a *d*-dimensional joint PDF, then according to chain rule, joint PDF can be written as

$$f(x_1, x_2, ..., x_d) = f_d(x_d) \prod_{k=d-1}^{1} f_{k|(k+1)...d}(x_k | x_{k+1}, ..., x_d) \tag{4}$$

Now, let's focus on the right-hand side of Eq. (4). According to the general product rule, the second term can be written as

$$\begin{aligned} f_{d-1|d}\left(x_{d-1}|x_d\right) &= \frac{f_{(d-1)d}\left(x_{d-1},x_d\right)}{f_d\left(x_d\right)} \\ &= \frac{c_{(d-1)d}\left\{F_{d-1}\left(x_{d-1}\right),F_d\left(x_d\right)\right\} \cdot f_{d-1}\left(x_{d-1}\right)f_d\left(x_d\right)}{f_d\left(x_d\right)} \\ &= c_{(d-1)d}\left\{F_{d-1}\left(x_{d-1}\right),F_d\left(x_d\right)\right\} \cdot f_{d-1}\left(x_{d-1}\right) \end{aligned} \quad (5)$$

Similarly, the third term can be written as

$$\begin{aligned} f_{d-2|(d-1)d}\left(x_{d-2}|x_{d-1},x_d\right) &= \frac{f_{(d-2)d|(d-1)}}{f_{d|d-1}} = \frac{c_{(d-2)d|(d-1)} \cdot f_{d-2|d-1} \cdot f_{d|d-1}}{f_{d|d-1}} \\ &= c_{(d-2)d|(d-1)} \cdot \frac{f_{(d-2)(d-1)}}{f_{d-1}} = c_{(d-2)d|(d-1)} \cdot c_{(d-2)(d-1)} \cdot f_{d-2} \end{aligned} \quad (6)$$

On the above decomposition, the terms $x_i$, $x_{i|j}$, $F_i(x_i)$ and $F_{i|j}(x_i|x_j)$ are neglected for the sake of simplicity.

Accordingly, the joint PDF $f(x_1,x_2,...,x_d)$ can be finally expressed as a product of marginal PDF and pair-copula density, which is consist of unconditional or conditional copula.

$$f(x_1,x_2,...,x_d) = \prod_{k=1}^{d} f_k(x_k) \prod_{j=1}^{d-1} \prod_{i}^{d-j} c_{i,i+j|i+1,...,i+j-1}\left\{F(x_i|x_{i+1},...,x_{i+j-1}),F(x_{i+j}|x_{i+1},...,x_{i+j-1})\right\} \quad (7)$$

The subscripts of $F(x_i|x_{i+1},...,x_{i+j-1})$ and $F(x_{i+j}|x_{i+1},...,x_{i+j-1})$ is neglected for the sake of simplicity.

To illustrate the construction of D-vine copula more clearly, a graphical model is introduced as shown in figure 2. It shows a construction of 5 variables D-vine copula. The graphical model has 4 trees and each edge of the vine is associated with the bivariate or conditional bivariate shown adjacent to the edge. For the tree $T_1$, each edge of vine is associated with the random variables of the bottom level. Then, tree $T_2$ is formed by the associated edge of $T_1$. Without loss of generality, the whole tree can be constructed. Finally, there will be a vine constructed by all the random variables. The above method of connecting two vine or random variables into a new vine is called pair-copula construction (PCC). The construction of pair copula is based on a symmetric difference method. This method is defined by $X \Delta Y := (X \setminus Y) \cup (Y \setminus X)$. For instance, symmetric difference of two set {1} and {2} is {12}, while {12} and {23} become {13 | 2}. As for a conditional one, for example, symmetric

difference of {13 | 2} and {24 | 3} is {14 | 23}.

## 2.2 One-step Bayesian copula model selection

In the past decades, various authors proposed different approaches of copula model selection. AIC, GOF and Bayesian method are three major methods. Recently, model selection based on Bayesian method is increasingly popular due to its feasibility and flexibility. For the traditional Bayesian method, it infers the copula function between input variables in two steps as shown in figure 3. Identify the marginal PDF of input variables first and then identify copula function between them. Two steps selection would inevitably lead to problems. For instance, when the input information is limited, the accuracy of marginal CDF identification might be in a poor level and some of the input information may lose in this process. Then the poor identification information from first step delivers to the copula identification step and would deteriorate the final result.

To address this problem, a novel one-step Bayesian copula model selection (OBCS) is proposed as shown in figure 3. The OBCS joins the marginal CDFs identification and copula function identification in one step, and thus, the error propagation can be avoided. In addition, it can make a full use of the input information and thus avoid information lost.

The OBCS method defines a candidate pool that contains different marginal CDFs and copula function combination, for instance, if both marginal CDFs are gauss and the copula function is Clayton, the combination would be Gauss-Clayton-Gauss. The combination contained in the pool can also be named as candidate. Therefore, various candidates will be included in the pool such as Gauss-Clayton-Gauss, Gama-Clayton-Gauss and Lognormal-Frank-Gama etc. as shown in figure 3. It can be easily derived that the number $N$ of candidates contained in the pool can be calculated by

$$N = n(n-1)m \tag{8}$$

where $n$ denotes number of candidate marginal CDF and $m$ denotes number of candidate copula function.

According to Bayesian theory, the realization possibility of each candidate in the pool can be given as

$$\Pr(h_l|D,I) = \frac{\Pr(D|h_l,I) \cdot \Pr(h_l|I)}{\Pr(D|I)} \tag{9}$$

where $h_l$ denotes data came from the candidate $M_l$, $D$ denotes input data, $I$ denotes additional information, $\Pr(h_l|D,I)$ denotes the possibility of $M_l$ under condition of $D$ and $I$, $\Pr(D|h_l,I)$ denotes a likelihood function, $\Pr(h_l|I)$ denotes prior information of candidate $M_l$ and $\Pr(D|I)$ denotes a normalized constant.

The candidate $M_l$ is consist of three parts, two marginal CDFs and a copula function. It is supposed that $f_{m_1}(x_1)$ and $f_{m_2}(x_2)$ denote two marginal PDFs, $u_i, v_i$ denotes two marginal CDFs, $c(u_i, v_i|\theta)$ denotes the copula function. Then the likelihood function of Eq. (9) can be expressed as

$$\Pr(D|h_l,I) = \Pr(D|h_l,\beta_1,\beta_2,I) = \prod_{i=1}^{N} c(u_i,v_i|\theta) \cdot f_{m_1}(x_1|\beta_1) \cdot f_{m_2}(x_2|\beta_2) \tag{10}$$

where $\theta$ represents parameter of the copula function and $\beta_1$, $\beta_2$ represents parameter of marginal PDFs. For instance, $\beta$ contains mean $\mu$ and standard deviation $\sigma$ when the marginal PDF is Gauss.

According to Eq. (10), Eq. (9) can be rewritten as

$$\begin{aligned} \Pr(h_l|D,I) &= \iiint \Pr(h_l,\beta_1,\beta_2,\theta|D,I) d\beta_1 d\beta_2 d\theta \\ &= \iiint \frac{\Pr(D|h_l,\beta_1,\beta_2,\theta,I)\Pr(\beta_1|h_l,I)\Pr(\beta_2|h_l,I)\Pr(\theta|h_l,I)\Pr(h_l|I)}{\Pr(D|I)} d\beta_1 d\beta_2 d\theta \end{aligned} \tag{11}$$

where $\Pr(\beta_1|h_l,I)$, $\Pr(\beta_2|h_l,I)$ and $\Pr(\theta|h_l,I)$ denote prior information of parameter $\beta_1$, $\beta_2$ and $\theta$ respectively.

Commonly, the additional information $I$ is defined as follow:

$I_1$: the candidate is equally probable when there is no other information given.

$I_2$: for a certain candidate, $\beta_1$, $\beta_2$ and $\theta$ are belong to the space of $\Lambda^{\beta_1}$, $\Lambda^{\beta_2}$ and $\Lambda^{\theta}$ respectively. Moreover, $\beta_1$, $\beta_2$ and $\theta$ is distributed uniformly when there is no other prior information.

According to the additional information $I_2$, prior of $\beta_1$, $\beta_2$ and $\theta$ can be expressed as

$$\Pr(\beta_1|h_l,I) = \begin{cases} \dfrac{1}{L(\Lambda^{\beta_1})} & \beta_1 \in \Lambda^{\beta_1} \\ 0 & \beta_1 \notin \Lambda^{\beta_1} \end{cases} \tag{12}$$

$$\Pr(\beta_2|h_l,I) = \begin{cases} \dfrac{1}{L(\Lambda^{\beta_2})} & \beta_2 \in \Lambda^{\beta_2} \\ 0 & \beta_2 \notin \Lambda^{\beta_2} \end{cases} \tag{13}$$

$$\Pr(\theta|h_l,I) = \begin{cases} \dfrac{1}{L(\Lambda^{\theta})} & \theta \in \Lambda^{\theta} \\ 0 & \theta \notin \Lambda^{\theta} \end{cases} \tag{14}$$

where $L(.)$ denotes the Lebesgue measure and is defined as the length of $\Lambda$.

According to the additional information $I_1$, the prior of the candidate can be expressed as

$$\Pr(h_l|I) = \frac{1}{N} \tag{15}$$

Substituting Eq. (15) and Eq. (12) to Eq. (14) into Eq. (11) gives the expression of

$$\Pr(h_l|D,I) = \iiint \frac{\prod_{i=1}^{N} c(u_i,v_i|\theta) \cdot f_{m_1}(x_{1i}|\beta_1) \cdot f_{m_2}(x_{2i}|\beta_2)}{N \Pr(D|I) L(\Lambda^{\beta_1}) L(\Lambda^{\beta_2}) L(\Lambda^{\theta})} d\beta_1 d\beta_2 d\theta \tag{16}$$

Since $\Pr(D|I)$ is a constant, we eliminate it and then the weight $w_l$ is given as

$$w_l = \iiint \frac{\prod_{i=1}^{N} c(u_i,v_i|\theta) \cdot f_{m_1}(x_{1i}|\beta_1) \cdot f_{m_2}(x_{2i}|\beta_2)}{N L(\Lambda^{\beta_1}) L(\Lambda^{\beta_2}) L(\Lambda^{\theta})} d\beta_1 d\beta_2 d\theta \tag{17}$$

Then the normalized weight $W_l$ can be expressed as

$$W_l = \frac{w_l}{\sum_{i}^{N} w_i} \tag{18}$$

where the subscript of $l$ denotes the $l^{th}$ candidate in the pool. The most likely candidate would be the one with the largest normalized weight $W_l$.

## 2.3 D-vine simulation

After identifying the types of marginal CDF and copula function, it is significant to generate samples for MCS. The D-vine simulation of $n$ dependent uniform [0, 1] samples is given. Firstly, sampling $n$ independent uniform [0, 1] random variables $u_1, u_2, ..., u_n$. Then the $n$ dependent uniform sample $x_1, x_2, ..., x_n$ can be obtained as

$$\begin{aligned} x_1 &= u_1 \\ x_2 &= F^{-1}(u_2 | x_1) \\ x_3 &= F^{-1}(u_3 | x_1, x_2) \\ &\vdots \\ x_n &= F^{-1}(u_n | x_1, x_2, ..., x_n) \end{aligned} \quad (19)$$

where $F^{-1}(.)$ denotes inverse distribution function. According to Joe. H [28], the conditional marginal CDF of the form $F_{x|\mathbf{v}}(x|\mathbf{v})$ can be calculated by

$$F_{x|\mathbf{v}}(x|\mathbf{v}) = \frac{\partial C_{xv_l|\mathbf{v}^{-1}}\{F_{x|\mathbf{v}^{-1}}(x|\mathbf{v}^{-1}), F_{v_l|\mathbf{v}^{-1}}(v_l|\mathbf{v}^{-1})\}}{\partial F_{v_l|\mathbf{v}^{-1}}(v_l|\mathbf{v}^{-1})} \quad (20)$$

where $v_l$ denotes the last component of vector $\mathbf{v}$.

For a bivariate case, Eq. (20) has the expression of

$$F_{x|v}(x|v) = \frac{\partial C_{xv}\{F_x(x), F_v(v)\}}{\partial F_v(v)} \quad (21)$$

Specifically, when the distribution of random variables $x$ and $v$ is uniform which indicates that $F(x) = x$ and $F(v) = v$. Then $g(x, v, \theta)$ can be defined to describe the conditional CDF of $F_{x|v}(x|v)$

$$g(x, v, \theta) = F_{x|v}(x|v) = \frac{\partial C_{xv}\{x, v, \theta\}}{\partial v} \quad (22)$$

According to Eq. (19), Eq. (21) and Eq. (22), the iteration algorithm can be obtained. The step of simulation is shown as follow:

*Step 1*: Sample $n$ independent uniform variables $u_1, u_2, ..., u_n$ from [0, 1].

*Step 2*: Let $x_1 = u_1$.

*Step 3*: Given $C_{12}, F_1$ and $F_2$, according to Eq. (21), $F_{2|1} = \frac{\partial C_{12}\{F_1, F_2, \theta\}}{\partial F_1}$, then $x_2$ can be sampled given $x_1$ and $u_2$.

*Step 4*: Given $C_{12}, F_1$ and $F_2$, $F_{1|2} = \frac{\partial C_{12}\{F_1, F_2, \theta\}}{\partial F_2}$ can be derived; given $C_{23}, F_2$ and $F_3$, $F_{3|2} = \frac{\partial C_{23}\{F_2, F_3, \theta\}}{\partial F_2}$ can be derived; given $C_{13|2}, F_{1|2}$ and $F_{3|2}$, $F_{3|12} = \frac{\partial C_{13|2}\{F_{1|2}, F_{3|2}, \theta\}}{\partial F_{1|2}}$ can be derived. Then $x_3$ can be sampled given $x_1, x_2$ and $u_3$.

…

Based on the above steps, an iteration algorithm is shown in algorithm 1. The algorithm consists of three closed-loops. For the first loop is used for the number needed to sample. The second and third loops are used to calculate the necessary conditional distribution functions. Moreover, $\theta_{i,j}$ indicates the parameters of corresponding copula density and *g*-function is the one defined by Eq. (22).

**Algorithm 1** D-vine copula simulation

Sample *n* independent uniform variables $u_1, u_2, ..., u_n$ in [0,1]

$x_1 = v_{1,1} = u_1$

$x_2 = v_{2,1} = g^{-1}(u_2, v_{1,1}, \theta_{1,1})$

$v_{2,2} = g(v_{2,1}, v_{1,1}, \theta_{1,1})$

Loop $i = 1, 2, ..., n$

$v_{i,1} = u_i$

    Loop $k = i-1, i-2, ..., 2$

        $v_{i,1} = g^{-1}(v_{i,1}, v_{i-1,2k-2}, \theta_{k,i-k})$

    End loop

$v_{i,1} = g^{-1}(v_{i,1}, v_{i-1,1}, \theta_{1,i-1})$

$x_i = v_{i,1}$

If $i == n$

    Stop

End if

$v_{i,2} = g(v_{i-1,1}, v_{i,1}, \theta_{1,i-1})$

$v_{i,3} = g(v_{i,1}, v_{i-1,1}, \theta_{1,i-1})$

If $i > 3$

    Loop $j = 2, 3, ..., i-2$

        $v_{i,2j} = g(v_{i-1,2j-2}, v_{i,2j-1}, \theta_{j,i-j})$

        $v_{i,2j+1} = g(v_{i,2j-1}, v_{i-1,2j-2}, \theta_{j,i-j})$

    End loop

End if

$v_{i,2i-2} = g(v_{i-1,2i-4}, v_{i,2i-3}, \theta_{i-1,1})$

End Loop

## 3. Evaluation methods for variable-stiffness composite

FEA is one of the most popular and flexible method in the engineering field. However,

evaluation time may be tremendous when large number of iterations is needed in MCS and in order to achieve a higher efficiency, two fast evaluation methods, reanalysis based FEA and BPNN model is introduced in this section. Theoretically, Reanalysis based FEA would show a higher efficient than FEA in iteration process. However, result shows that the although the efficiency is improved, time consumption is still relatively large. To this end, a BPNN surrogated model is proposed to significantly improve the efficiency of the whole program.

## 3.1 Reanalysis assisted FEA

In this study, FE modeling of VS composite laminate is based on the Mindlin theory. Accordingly, the DOF of nodes is defined as $d_e = [\mu, v, \omega, \theta_x, \theta_y]$. where $\mu, v, \omega$ denotes the mid-plane displacement in *x, y* and *z* direction respectively. $\theta_x$ and $\theta_y$ represent the rotation angle toward x-axis and y-axis respectively.

According to Mindlin plate theory, the element contains three types of working condition, in-plane condition, bending condition and shear condition. Thus, element stiffness matrix can be given by

$$k_e = k_m + k_b + k_s \tag{23}$$

where $k_e$ denotes the element stiffness matrix and $k_m$, $k_b$ and $k_s$ denote in-plane, bending and shearing stiffness matrix respectively, and they can be expressed as

$$k_m = \int_\Omega B_m^T D_m B_m d\Omega \tag{24}$$

$$k_b = \int_\Omega B_b^T D_b B_b d\Omega \tag{25}$$

$$k_s = \int_\Omega B_s^T D_s B_s d\Omega \tag{26}$$

where $B_m$, $B_b$ and $B_s$ denote the corresponding strain matrices and $D_m$, $D_b$ and $D_s$ denote the corresponding constitutive matrices. Calculation is based on the local coordinate of the fiber angle orientation while the definition of fiber angle is always on the global coordinate. Then, the relation between local and global coordinate can be express as

$$\theta_L = \theta_G - \theta_T \tag{27}$$

where $\theta_L$ and $\theta_G$ denote local and global coordinate respectively and $\theta_T$ denotes the difference

between global and local angles

After introducing the transformation matrix, Eq. (24) to Eq. (26) can be expressed as

$$k_m = \int_\Omega B_m^T T_m^T D_m T_m B_m d\Omega \tag{28}$$

$$k_b = \int_\Omega B_b^T T_b^T D_b T_b B_b d\Omega \tag{29}$$

$$k_s = \int_\Omega B_s^T T_s^T D_s T_s B_s d\Omega \tag{30}$$

where $T_m$, $T_b$ and $T_s$ denote corresponding transformation matrices and have the expression of

$$T_m = T_b = \begin{bmatrix} \cos^2\theta_L & \sin^2\theta_L & \cos\theta_L \sin\theta_L \\ \sin^2\theta_L & \cos^2\theta_L & -\cos\theta_L \sin\theta_L \\ -2\cos\theta_L \sin\theta_L & 2\cos\theta_L \sin\theta_L & \cos^2\theta_L - \sin^2\theta_L \end{bmatrix}$$

$$T_s = \begin{bmatrix} \cos\theta_L & \sin\theta_L \\ \sin\theta_L & \cos\theta_L \end{bmatrix}$$

and

$$D_m = D_b = \begin{bmatrix} Q_{11} & Q_{12} & 0 \\ Q_{12} & Q_{22} & 0 \\ 0 & 0 & Q_{66} \end{bmatrix}$$

$$D_s = \begin{bmatrix} Q_{44} & 0 \\ 0 & Q_{55} \end{bmatrix}$$

where $Q_{11}$, $Q_{12}$, $Q_{22}$, $Q_{44}$, $Q_{55}$ and $Q_{66}$ have the expression of

$$\begin{cases} Q_{11} = \dfrac{E_L}{1-\nu_{LT}\nu_{TL}} \\ Q_{22} = \dfrac{E_T}{1-\nu_{LT}\nu_{TL}} \\ Q_{12} = \dfrac{\nu_{LT} E_L}{1-\nu_{LT}\nu_{TL}} \\ Q_{44} = G_{TN} \quad Q_{55} = G_{LN} \quad Q_{66} = G_{LT} \end{cases} \tag{31}$$

where $E_L$ and $E_T$ denote the longitudinal and transverse elasticity modulus; $G_{TN}$, $G_{LT}$ and $G_{LN}$ denote in-plane and out-of-plane shear modulus respectively; $\nu_{LT}$ and $\nu_{TL}$ denote major and minor Poisson ratios.

It is supposed that the total ply of plate is $n_p$ and thickness of each ply is $t_i (i=1,2,...,n_p)$. Moreover, it is assumed that the coordinate in z direction for each ply is $z_i$ and the origin of the coordinate is set at the mid-plane as shown in figure 4. Then the relationship between $t_i$

and $z_i$ can be expressed as

$$t_i = z_i - z_{i-1} \tag{32}$$

where $z_i$ denotes the coordinate of upper plane of each ply and $z_{i-1}$ denotes the coordinate of bottom plane of each ply.

Considering the plate thickness, the element stiffness can be rewritten as

$$k_m = \sum_{i=1}^{n_p} \int_{z_{i-1}}^{z_i} \int_A B_m^T T_m^{i^T} D_m T_m^i B_m dA dz = \sum_{i=1}^{n_p} t_i \int_A B_m^T T_m^{i^T} D_m T_m^i B_m dA \tag{33}$$

$$k_b = \sum_{i=1}^{n_p} \int_{z_{i-1}}^{z_i} z^2 \int_A \bar{B}_b^T T_b^{i^T} D_b T_b^i \bar{B}_b dA dz = \sum_{i=1}^{n_p} \frac{z_i^3 - z_{i-1}^3}{3} \int_A \bar{B}_b^T T_b^{i^T} D_b T_b^i \bar{B}_b dA \tag{34}$$

$$k_s = \sum_{i=1}^{n_p} \int_{z_{i-1}}^{z_i} \int_A B_s^T T_s^{i^T} D_s T_s^i B_s dA dz = \sum_{i=1}^{n_p} t_i \int_A B_s^T T_s^{i^T} D_s T_s^i B_s dA \tag{35}$$

where $T_m^i$, $T_b^i$ and $T_s^i$ denote the transformation matrix for each ply; $\bar{B}_b$ denotes curvature matrix and has the expression of

$$B_b = -z\bar{B}_b \tag{36}$$

For the VS composite laminate, the element stiffness matrix has a relation with the fiber angle orientation. Therefore, the element stiffness matrix can be written as

$$\begin{aligned} k &= k(\theta_L) = k_m(\theta_L) + k_b(\theta_L) + k_s(\theta_L) \\ &= \sum_{i=1}^{n_p} t_i \int_A B_m^T T_m^{i^T}(\theta_L) D_m T_m^i(\theta_L) B_m dA + \sum_{i=1}^{n_p} \frac{z_i^3 - z_{i-1}^3}{3} \int_A \bar{B}_b^T T_b^{i^T}(\theta_L) D_b T_b^i(\theta_L) \bar{B}_b dA + \sum_{i=1}^{n_p} t_i \int_A B_s^T T_s^{i^T}(\theta_L) D_s T_s^i(\theta_L) B_s dA \\ &= \sum_{i=1}^{n_p} (t_i \int_A B_m^T \bar{D}_m^i(\theta_L) B_m dA + \frac{z_i^3 - z_{i-1}^3}{3} \int_A \bar{B}_b^T \bar{D}_b^i(\theta_L) \bar{B}_b dA + t_i \int_A B_s^T \bar{D}_s^i(\theta_L) B_s dA) \end{aligned} \tag{37}$$

where

$$\begin{cases} \bar{D}_m^i(\theta_L) = T_m^{i^T}(\theta_L) D_m T_m^i(\theta_L) \\ \bar{D}_b^i(\theta_L) = T_b^{i^T}(\theta_L) D_b T_b^i(\theta_L) \\ \bar{D}_s^i(\theta_L) = T_s^{i^T}(\theta_L) D_s T_s^i(\theta_L) \end{cases} \tag{38}$$

Considering the fiber angle deviation, the local fiber angle can be rewritten as

$$\hat{\theta}_L = \theta_L + \varepsilon \tag{39}$$

where $\varepsilon$ denotes the fiber angle deviation of each ply. Then the global stiffness matrix can be written as

$$K = K(\hat{\theta}_L) = \bigcup_e \sum_{i=1}^{n_p} (t_i \int_A B_m^T \bar{D}_m^i(\hat{\theta}_L) B_m dA + \frac{z_i^3 - z_{i-1}^3}{3} \int_A \bar{B}_b^T \bar{D}_b^i(\hat{\theta}_L) \bar{B}_b dA + t_i \int_A B_s^T \bar{D}_s^i(\hat{\theta}_L) B_s dA) \tag{40}$$

where

$$\begin{cases} \bar{D}_m^i(\hat{\theta}_L) = T_m^{i^T}(\hat{\theta}_L) D_m T_m^i(\hat{\theta}_L) \\ \bar{D}_b^i(\hat{\theta}_L) = T_b^{i^T}(\hat{\theta}_L) D_b T_b^i(\hat{\theta}_L) \\ \bar{D}_s^i(\hat{\theta}_L) = T_s^{i^T}(\hat{\theta}_L) D_s T_s^i(\hat{\theta}_L) \end{cases} \tag{41}$$

then let $r_0$ denotes the initial displacement and it can be expressed as

$$r_0 = K_0^{-1} R \tag{42}$$

where $K_0 = K_0(\theta_L)$ denotes the initial stiffness matrix and $R$ denotes load vector.

For each iterative step, the deviation of fiber angle will be added to the ideal fiber angle, which lead to a change of the initial stiffness matrix. Then the equilibrium equation after updating is given by

$$Kr = R \tag{43}$$

where $R$ stay unchanged in this study and $K = K(\hat{\theta}_L)$ denotes a new stiffness matrix and can be expressed as

$$K = K_0 + \Delta K \tag{44}$$

where $\Delta K$ denotes the change part of $K$. Substituting Eq. (44) into Eq. (43) gives

$$(K_0 + \Delta K) r = R \tag{45}$$

then displacement $r$ can be expressed as

$$r = (K_0 + \Delta K)^{-1} R = (1+B)^{-1} K_0^{-1} R = (1+B)^{-1} r_0 \tag{46}$$

where $I$ denote a unit matrix and

$$B = K_0^{-1} \Delta K \tag{47}$$

Then applying binomial series to Eq. (46) gives

$$r = r_0 - B r_0 + B^2 r_0 + ... \tag{48}$$

The displacement vector $r$ can be approximated by a linearly independent basis vectors of $s$. Then the displacement vector $r$ becomes

$$r = y_1 r_1 + y_2 r_2 + ... + y_s r_s = r_B y \tag{49}$$

where $r_B = [r_1, r_2, ..., r_s]$ and $y = [y_1, y_2, ..., y_s]^T$. Comparing Eq. (48) and Eq. (49) vector $r_B$ can

be constructed as

$$r_1 = r_0 \tag{50}$$

$$r_i = -Br_{i-1} \quad i = 2,3,...,s \tag{51}$$

Substituting Eq. (49) into Eq. (42) and multiply both side of the equation with $r_B$ gives

$$K_r y = R_r \tag{52}$$

where

$$K_r = r_B^T K r_B \tag{53}$$

$$R_r = r_B^T R \tag{54}$$

After solving Eq. (52), $y$ can be derived and substituted it into Eq. (49), the displacement $r$ can be obtained.

## 3.2 BPNN surrogate model

Although the calculation of MCS using reanalysis method is feasible, the computational cost is still relatively huge. To this end, BPNN is presented herein to improve the efficiency of the entire MCS further.

BPNN is a multilayer feedforward neural network consisted of input layer, hidden layer and output layer. It has a strong ability of target identification and regression. Moreover, BPNN might be the one of methods that maps a highly nonlinear between input and output data. The structure of BPNN is shown in figure 5.

Mathematically, the BPNN has the expression of

$$f_k(\boldsymbol{T}) = \sum_{j=1}^{m} \omega_{jk} f(\sum_{i=1}^{n} \omega_{ij} x_i(\boldsymbol{T}) + b_i) + b_j \tag{55}$$

where $\boldsymbol{T}$ denotes the vector of input data, $\omega_{ij}$ and $b_i$ denote the weighted and threshold from the input layer to the hidden layer, $\omega_{jk}$ and $b_j$ denote the weighted and threshold from the hidden layer to the output layer, $f(.)$ denotes Tan-Sigmoid transfer function and has the expression of

$$f(x) = \frac{1-e^{-x}}{1+e^{-x}} \tag{56}$$

In this study, BPNN is used to serve as a surrogate model in the MCS. According to Hornik *et al.*[25], one hidden layer is capable for approximation when the hidden unit is sufficient. Moreover, 34 hidden units is chosen in the hidden layer and evidence will be covered in the numerical example. Therefore, the BPNN with one hidden layer and 34 hidden units is used in this strategy. Subsequently, based on the reanalysis method, the BPNN model can be trained.

However, it is significant to evaluate the accuracy of BPNN model. Mean relative accuracy is a criterion and has the expression of

$$acc = \frac{1}{t}\sum_{i=1}^{t} 1 - \frac{|y_i - \tilde{y}_i|}{\tilde{y}_i} \tag{57}$$

where $t$ denotes the number of testing samples and $y_i$, $\tilde{y}_i$ denotes the result generated from the BPNN model and reanalysis respectively. Furthermore, determination coefficient $R^2$ is another important coefficient to evaluate the BPNN model and has the expression of

$$R^2 = 1 - \frac{\sum_{i=1}^{t}(y_i - \tilde{y}_i)^2}{\sum_{i=1}^{t}(y_i - \bar{y}_i)^2} \tag{58}$$

where $\bar{y}_i$ denotes the mean value of result of reanalysis method. $R^2$ indicates how well the regression outcome can approximate the real data point. It provides a criterion of feasibility of the regression model with the range of [0, 1] and if it approaches 1, it means that the BPNN model is capable to replace the FE model.

Specifically, distribution of the result is used to evaluate the feasibility of BPNN model in this study. The distribution of result of BPNN model should be similar to the reanalysis result. In details, the mean value, variance and bandwidth of result of the BPNN model should be evaluated.

## 4. Numerical examples

### 4.1 Variable-stiffness composite hole plate

An 8-ply VS composite hole plate is investigated here to verify the feasibility of the suggest strategy. The geometry structure and material properties of the VS composite hole plate are shown respectively in figure 6 and table 1 [17].

The plate is fixed on the left side and a uniform force 2 *N/m* is applied on the right side along *X* direction. Moreover, quadratic are used as the path function of the fiber angle, which can be expressed as

$$z(x, y) = x + a_1 y + a_2 xy + a_3 x^2 + a_4 y^2 \tag{59}$$

The parameters of Eq. (59) can be determined by genetic algorithm according to [5] strategy and the result is shown in table 2.

According to the path function, the fiber path of the plate is shown in figure 7. Then the composite hole plate can be manufactured according to the path function. According to the OBCS method, the marginal probability distribution is identified to be Gauss and the copula function is Frank. The Frank copula has the expression of

$$C(\mu, \nu) = -\frac{1}{\theta} \ln(1 + \frac{(e^{-\theta \mu} - 1)(e^{-\theta \nu} - 1)}{e^{-\theta} - 1}) \quad 0 \leq \theta < \infty \tag{60}$$

More commonly, parameter $\theta$ can be expressed in a general way, namely Kendall's tau expression. Kendall's tau is a correlation parameter describing the linear relation between two random variables. The relation between Kendall's tau $\tau$ and Frank parameter $\theta$ can be expressed as

$$\tau = 1 - \frac{4}{\theta}(1 - \frac{1}{\theta}\int_0^\theta \frac{t}{e^t - 1} dt) \quad \tau \in [-1, 1] \tag{61}$$

Then, applying the D-vine simulation, a dependent sample of fiber angle deviation can be derived. During the simulation, the copula is chosen to be the Frank copula. The fiber angle is arranged to be symmetric, so there will be a close relation between ply $1^{st}$ and $2^{nd}$, $3^{rd}$ and $4^{th}$, $5^{th}$ and $6^{th}$, $7^{th}$ and $8^{th}$ respectively. Therefore, it is reasonable to suppose that the Kendall's tau equals to -0.7 for the above plies while the others Kendall's tau is set to be 0.3 while constructing PCC. Figure 8 shows the sample set distribution between plies and it can be indicated that angle deviation between ply $1^{st}$ and $2^{nd}$, $3^{rd}$ and $4^{th}$, $5^{th}$ and $6^{th}$, $7^{th}$ and $8^{th}$ show a relatively strong linear correlation but diverse between other plies. The total number of sample data set is 10000 and the data is transformed to a normal distribution of mean value $\mu = 0$ and standard deviation $\sigma = 0.2$ via inverse transform since the angle deviation would not be too large in practice.

Sequentially, based on the sample data set, distribution of displacement of the VS composite hole plate can be obtained through reanalysis method and BPNN model. A comparison of results is shown in figure 9. It can be seen that the distribution is highly resemble. The BPNN model has an almost the same shape as well as bandwidth compared to reanalysis method. The peak of histogram of BPNN model and the reanalysis both occur in an identical displacement, about $2.1 \times 10^{-6} m$ in *X* direction and about $5.8 \times 10^{-7} m$ in *Y* direction.

Table 3 shows the comparison of mean value, variance and bandwidth of these two methods. It can be seen that the mean value of these two models is very close, with the relative error of 0.03% in *X* direction and 0.18% in *Y* direction. Since the distribution of displacement is similar with a gauss distribution, a normal curve is used to fit the distribution and the result is shown in figure 9. Figure 10 shows the CDF of reanalysis method and BPNN model. It can be found that two curves are almost coincide in both direction. Moreover, figure 11 shows the convergence of MCS and the mean response of the hole plate. It can be seen from the figure that the mean response converges at about 8000 iteration times in *X* direction and about 5000 in *Y* direction.

Table 4 shows the relative accuracy and the $R^2$ of the BPNN model. The relative accuracy is in a high level, up to 99% for *X* direction and 92% for *Y* direction. Moreover, the $R^2$ is also relatively large, which means that the BPNN model can accurately surrogate the FE model.

## 4.2 Variable-stiffness composite plane beam

Another numerical example, a VS composite plane beam, which is identical to [5], is introduced here using the suggested strategy. It should be noticed that though the VS plane beam is identical to [5], the purpose of this strategy is focus on investigating the uncertainty of response while [5] devoted to optimization. The geometry of a 4 plies VS composite plane beam is shown in figure 12.

The plane beam is fixed at both side and a concentrated force $F = 100 mN$ is applied at the center of it. The material parameter of the plane beam is same as the hole plate as shown in table 1. In addition, the fiber path of the 1st and the 2nd ply, the 3rd and 4th ply is arranged to be symmetric and path function is cubic, with the expression of

$$z = x + a_1 y + a_2 xy + a_3 x^2 + a_4 y^2 + a_5 x^2 y + a_6 xy^2 + a_7 x^3 + a_8 y^3 \qquad (62)$$

Using the genetic algorithm, the parameter of path function Eq. (62) can be determined and the result is shown in table 5. Based on the path function, the fiber path can be calculated and the result is shown in figure 13.

Applying the OBCS method, the marginal CDFs of fiber orientation deviation is Gauss and the copula is also Frank and has the expression of Eq. (60). Similarly, Kendall's tau between the 1$^{st}$ and the 2$^{nd}$ ply, the 3$^{rd}$ and the 4$^{th}$ ply are set to be $\tau = -0.7$ and the others are set to be $\tau = 0.3$. Applying the above information into D-vine copula simulation, a dependent uniform fiber angle deviation can be obtained. Figure 14 shows the angle deviation distribution between plies. Consistently, ply 1$^{st}$ and 2$^{nd}$, 3$^{rd}$ and 4$^{th}$ exhibit a much higher linear correlation while others are diverse. The size of sample set is 10000 and identically, samples are transformed to a normal distribution of mean value $\mu = 0$ and standard deviation $\sigma = 0.2$ via inverse transform.

It can be found from numerical example 1 that BPNN model is sufficient accurate compared with the reanalysis assisted FEA of VS composite. Therefore, BPNN model based on reanalysis is directly used in this case. The distribution of displacement of the VS composite plane beam is shown in figure 15.

It can be seen from the figure that the distribution of displacement doesn't seem to be a Gauss while the random variable is Gauss. In other word, it seems to be no clue that the distribution of displacement conforms to the distribution of angle deviation of VS composite. Moreover, a lognormal curve is used to fit the displacement distribution and PDF and CDF are shown in figure 15 and figure 16. The mean value, variance and bandwidth of the distribution are presented in table 6. Moreover, the MCS convergence are also study and the result is shown in figure 17, along with its mean value response. It can be found from the figure that the maximum displacement converges at 4000 iteration times in *X* direction and 6000 iteration times in *Y* direction.

## 4.3 Comparison of efficiency and determination of hidden unit

Using the reanalysis based BPNN model, the efficiency of MCS is substantially improved.

Table 7 shows the comparison of efficiency of above two numerical examples using different methods. Compared with the complete-based evaluation, the efficiency of reanalysis has been significantly improved. Moreover, it can be found that the efficiency of MCS using BPNN model is higher than the reanalysis method. This is because although reanalysis method can improve the efficiency of the FEA, the computational cost of physical model (FEM model) is unavoidably larger than the mathematical model (BPNN model).

Moreover, to verify the hidden unit used in the BPNN model, figure 18 and 19 show the relation between hidden units and the relative accuracy as well as $R^2$. It can be found from figure 18 that the relative error quickly converges accurately as the hidden unit increased. However, the $R^2$ fluctuates as the hidden unit increased. When the number of hidden unit is 34, the $R^2$ is at a relatively high level. Therefore, a one-layer PNN model with 34 hidden unit is chosen in this study.

## 5. Conclusion

In this study, an uncertainty analysis method considering correlation random variables to VS composite structure is performed. The entire uncertainty analysis is based on the MCS. To address the problem of coupling of variables of VS composite, a D-vine copula based uncertainty analysis is introduced. A novel OBCS assisted D-vine simulation is introduced as the sampling method of the suggested strategy, which is capable of qualifying the coupling random variables. Firstly, the copula function as well as the marginal CDFs of fiber angle deviation is identified by using suggested OBCS. Sequentially, based on the above information, dependent uniform samples are generated through a D-vine simulation. Moreover, a fast solver, reanalysis algorithm is used to evaluate expensive forward problems. However, the efficiency of MCS is insufficient although involving the reanalysis method. Therefore, to enhance efficiency of reanalysis method, the BPNN is employed to construct surrogate model instead of reanalysis,

Two numerical examples are used to evaluate performance of suggested methods. It can be found that the suggested strategy can obtain the distribution of displacement of VS composite efficiently and accurately. Moreover, it can be concluded that fiber angle deviation has a great influence on VS composite performance and the distribution of the response may not conform to the distribution of the random variables. Finally, efficiency of the BPNN model and reanalysis method are also evaluated. The major contribution of this study can be summarized as follows:

(1) This study attempts to investigate the stiffness characteristic of VS composite when considering uncertainty.

(2) Random variables are assumed to be coupled in this strategy. To solve the coupling problem, D-vine copula is introduced. Moreover, on the basis of traditional Bayesian model selection method, a novel OBCS method is introduced, which is capable to verify the copula relation and marginal distribution of random variables in a one-step process.

(3) The reanalysis formulations for VS composite is deduced. Based on the reanalysis method, the distribution of deformation of VS composite can be obtained efficiently. Compared with the full analysis, the efficiency of reanalysis method based MSC can be further significantly improved.

(4) BPNN model is also introduced based on the reanalysis evaluated samples. Results suggest that the efficiency of BPNN based MCS is substantially improved and accuracy is also guaranteed.

## Acknowledgments

This work has been supported by Project of National Key R&D Program of China 2017YFB0203701, Program of National Natural Science Foundation of China under the Grant Numbers 11572120.

## References

[1]. Asmanoglo, Tobias, and A. Menzel. "A multi-field finite element approach for the modelling of fibre-reinforced composites with fibre-bending stiffness." Computer Methods in Applied Mechanics & Engineering 317(2017):1037-1067.
[2]. Justo, J., et al. "Characterization of 3D printed long fibre reinforced composites." Composite Structures 185(2017).
[3]. Wu, L, et al. "A multiscale mean-field homogenization method for fiber-reinforced composites with gradient-enhanced damage models." Computer Methods in Applied

Mechanics & Engineering 233-236.4(2012):164-179.

[4]. Setoodeh, Shahriar, Z. Gürdal, and L. T. Watson. "Design of variable-stiffness composite layers using cellular automata." Computer Methods in Applied Mechanics & Engineering 195.9–12(2006):836-851.

[5]. Huang, Guanxin, Hu Wang, and Guangyao Li. "An efficient reanalysis assisted optimization for variable-stiffness composite design by using path functions." Composite Structures 153 (2016): 409-420.

[6]. Hao, Peng, et al. "Isogeometric buckling analysis of composite variable-stiffness panels." Composite Structures 165(2017):192-208.

[7]. Hao, Peng, et al. "Buckling optimization of variable-stiffness composite panels based on flow field function." Composite Structures 181 (2017): 240-255.

[8]. Vanmarcke, E. M. S. G. I., et al. "Random fields and stochastic finite elements." Structural Safety 3.3-4 (1986): 143-166.

[9]. Park, J. S., C. G. Kim, and C. S. Hong. "Stochastic finite element method for laminated composite structures." Journal of reinforced plastics and composites 14.7 (1995): 675-693.

[10]. Cui, X. Y., X. B. Hu, and Y. Zeng. "A Copula-based perturbation stochastic method for fiber-reinforced composite structures with correlations." Computer Methods in Applied Mechanics and Engineering 322 (2017): 351-372.

[11]. Stefanou, George, and Manolis Papadrakakis. "Stochastic finite element analysis of shells with combined random material and geometric properties." Computer Methods in Applied Mechanics and Engineering 193.1 (2004): 139-160.

[12]. Chen, Nian-Zhong, and C. Guedes Soares. "Spectral stochastic finite element analysis for laminated composite plates." Computer Methods in Applied Mechanics and Engineering 197.51 (2008): 4830-4839.

[13]. Sepahvand, K. "Spectral stochastic finite element vibration analysis of fiber-reinforced composites with random fiber orientation." Composite Structures 145 (2016): 119-128.

[14]. Jeong, H. K., and R. A. Shenoi. "Probabilistic strength analysis of rectangular FRP plates using Monte Carlo simulation." Computers & Structures 76.1 (2000): 219-235.

[15]. Ghanem, Roger. "Hybrid stochastic finite elements and generalized Monte Carlo simulation." TRANSACTIONS-AMERICAN SOCIETY OF MECHANICAL ENGINEERS JOURNAL OF APPLIED MECHANICS 65 (1998): 1004-1009.

[16]. Frangopol, Dan M., and Sebastien Recek. "Reliability of fiber-reinforced composite laminate plates." Probabilistic Engineering Mechanics 18.2 (2003): 119-137.

[17]. Yang, Zeng, E. Li, and H. Wang. "Fast variable stiffness composite cylinder uncertainty analysis by using reanalysis assisted Copula function." 2016:10006.

[18]. Sklar, M. "Fonctions de Répartition À N Dimensions Et Leurs Marges." Publ.inst.statist.univ.paris 8(1959):229-231.

[19]. Huard, David, Guillaume Évin, and Anne-Catherine Favre. "Bayesian copula selection." Computational Statistics & Data Analysis 51.2 (2006): 809-822.

[20]. Bedford, Tim, and Roger M. Cooke. "Vines: A new graphical model for dependent random variables." Annals of Statistics (2002): 1031-1068.

[21]. Mourelatos, Zissimos P, and E. Nikolaidis. "An Efficient Re-Analysis Methodology for Vibration of Large-Scale Structures (PREPRINT)." International Journal of Vehicle

Design 61.1-4(2009):37.

[22]. Wang, Hu, et al. ""Seen Is Solution" a CAD/CAE integrated parallel reanalysis design system." Computer Methods in Applied Mechanics & Engineering 299(2016):187-214.

[23]. Wang, Hu, Enying Li, and Guangyao Li. "A parallel reanalysis method based on approximate inverse matrix for complex engineering problems." Journal of Mechanical Design 135.8 (2013): 081001.

[24]. Sherman, Jack. "Adjustment of an inverse matrix corresponding to changes in the elements of a given column or a given row of the original matrix." Annals of Mathematical Statistics 20.4 (1949): 621.

[25]. Hornik, Kurt, Maxwell Stinchcombe, and Halbert White. "Multilayer feedforward networks are universal approximators." Neural networks 2.5 (1989): 359-366.

[26]. García-Iruela, Alberto, et al. "Comparison of modelling using regression techniques and an artificial neural network for obtaining the static modulus of elasticity of Pinus radiata D. Don. timber by ultrasound." Composites Part B: Engineering 96 (2016): 112-118.

[27]. Aas, Kjersti, et al. "Pair-copula constructions of multiple dependence." Insurance: Mathematics and economics 44.2 (2009): 182-198.

[28]. Joe, Harry. "Families of m-variate distributions with given margins and m (m-1)/2 bivariate dependence parameters." Lecture Notes-Monograph Series (1996): 120-141.

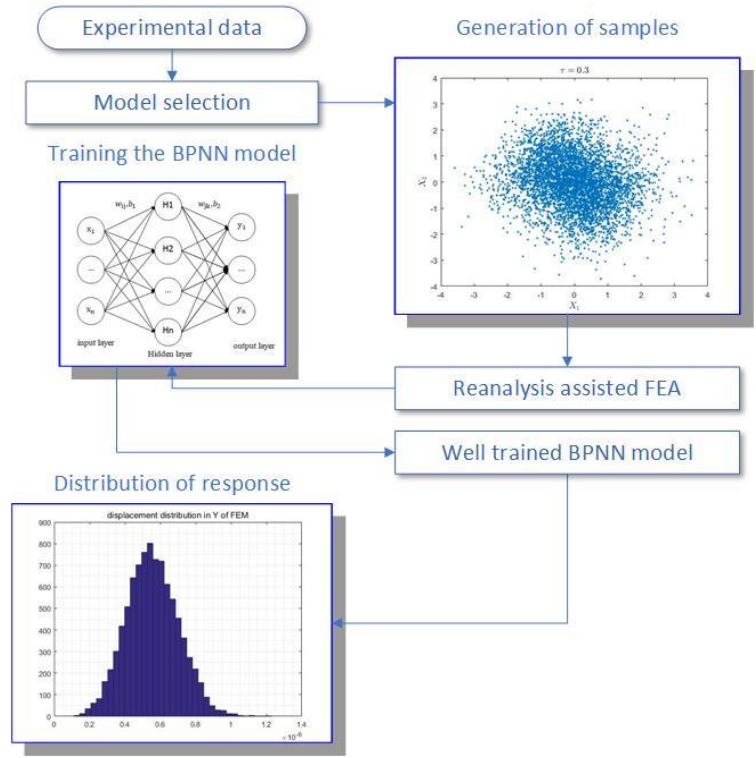

Figure 1. Framework of D-vine copula-based coupling uncertainty analysis

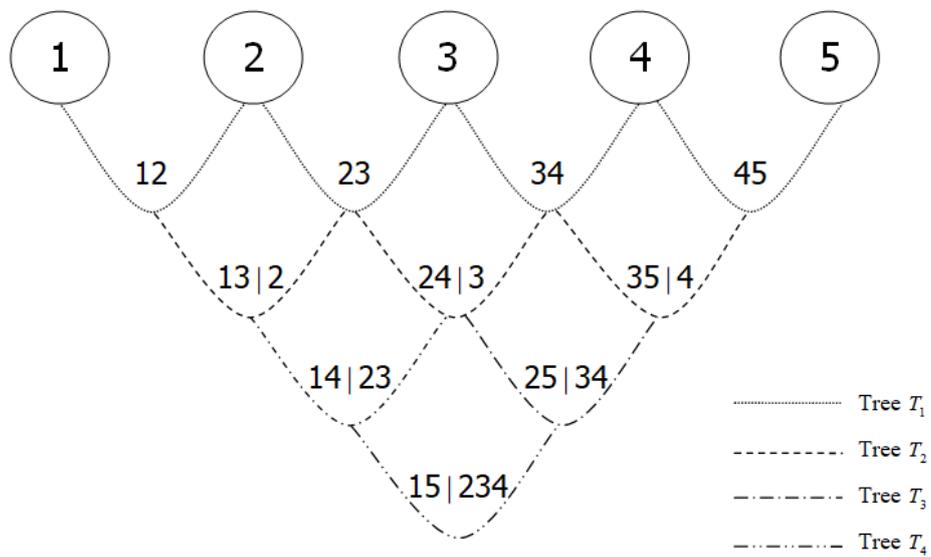

Figure 2. D-vine tree representation

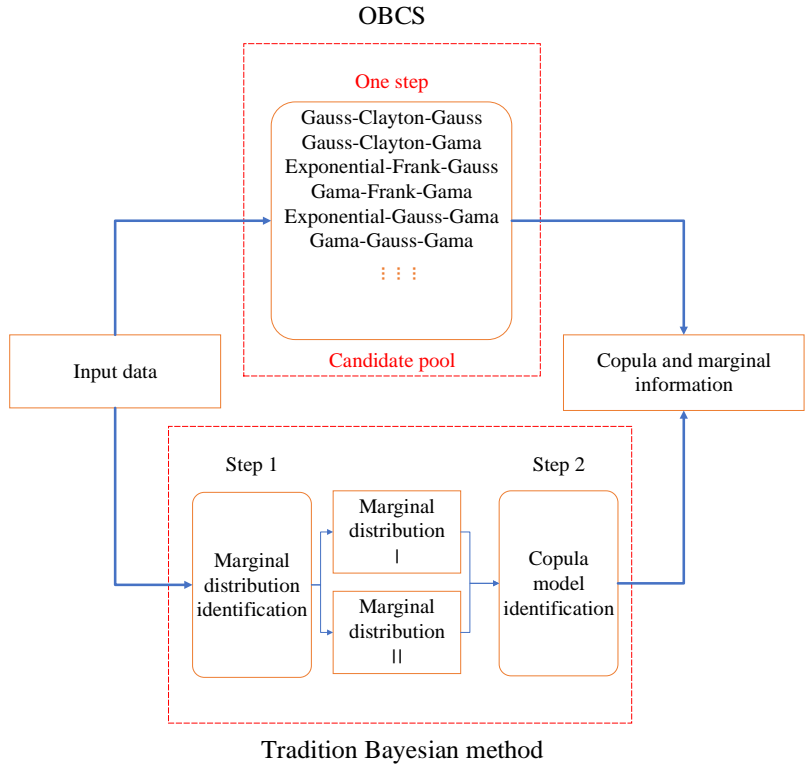

Figure 3. Comparison of tradition Bayesian and OBCS method

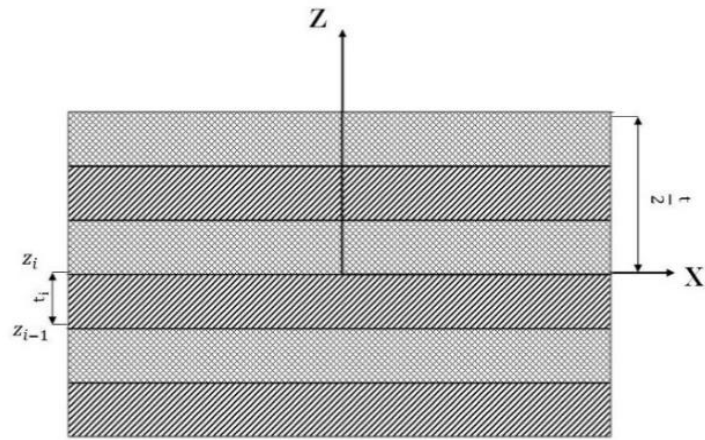

Figure 4. Coordinate system of laminate

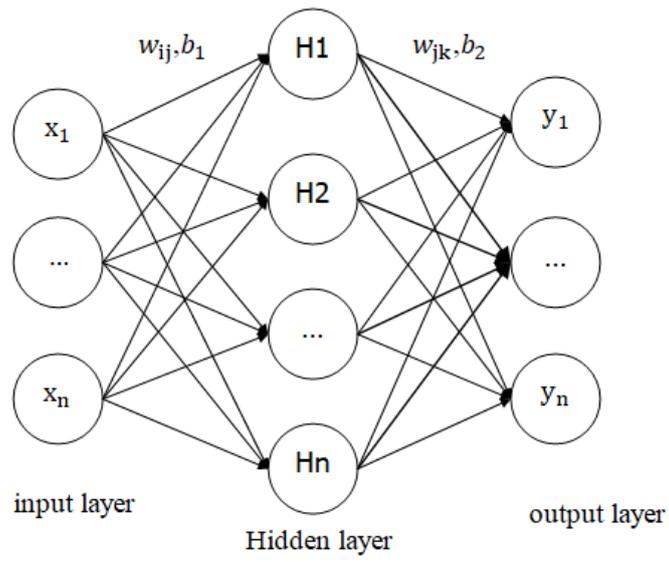

Figure 5. Structure of BPNN

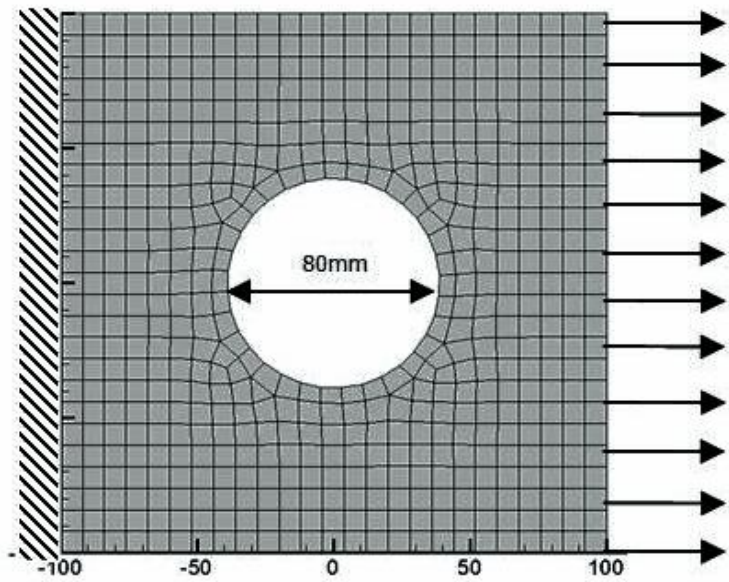

Figure 6. Geometry and boundary condition of hole plate

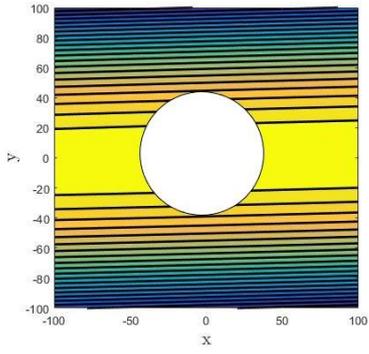
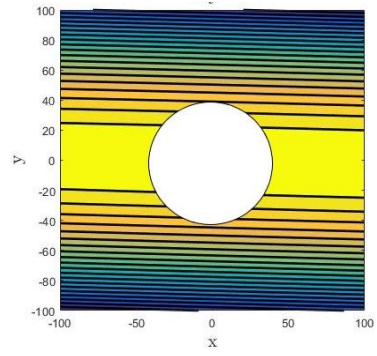

Ply 1                                    Ply 2

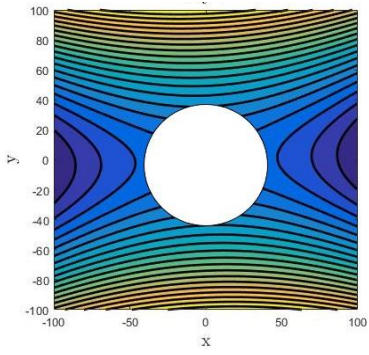
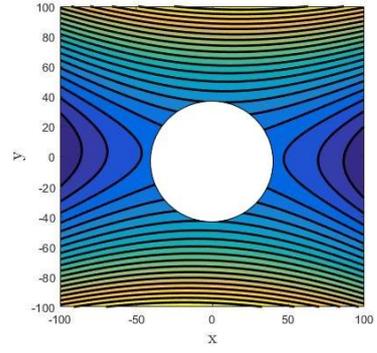

Ply 3                                    Ply 4

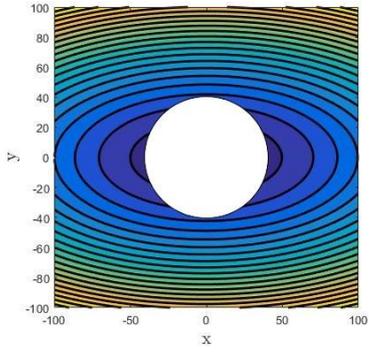
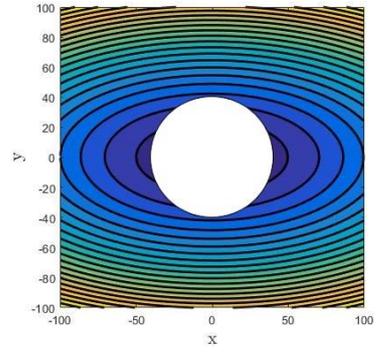

Ply 5                                    Ply 6

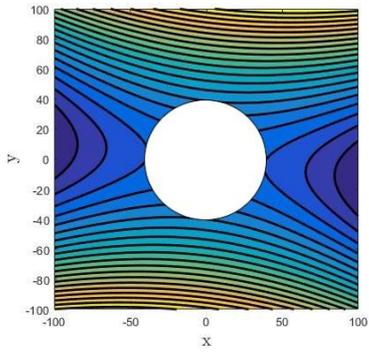
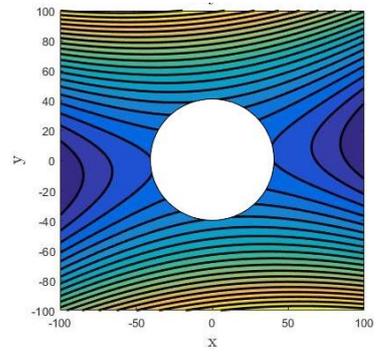

Ply 7                      Ply 8

Figure 7 Fiber path of hole plate

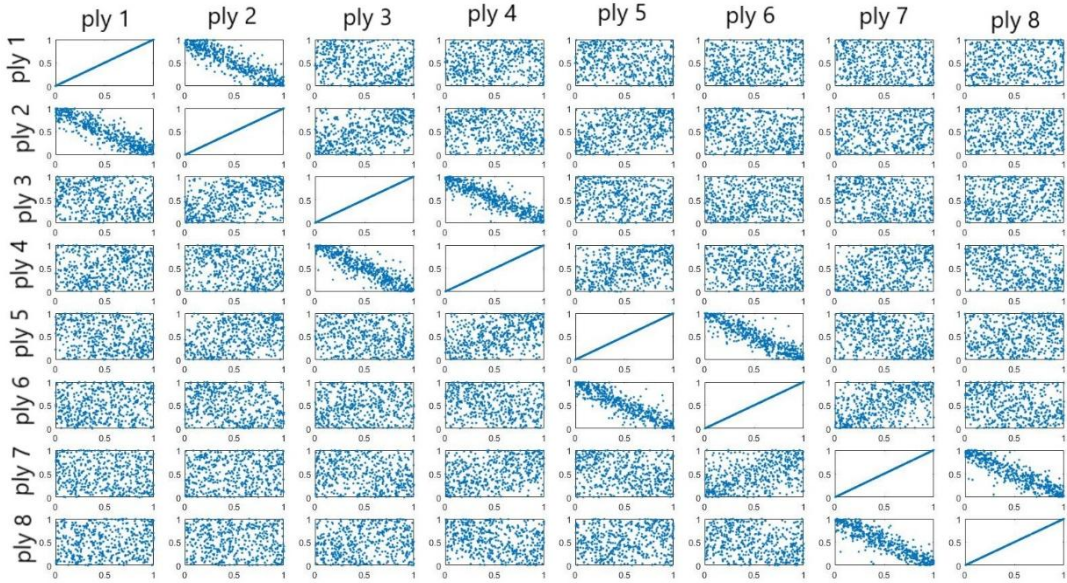

Figure 8. Sample distribution of numerical example 1

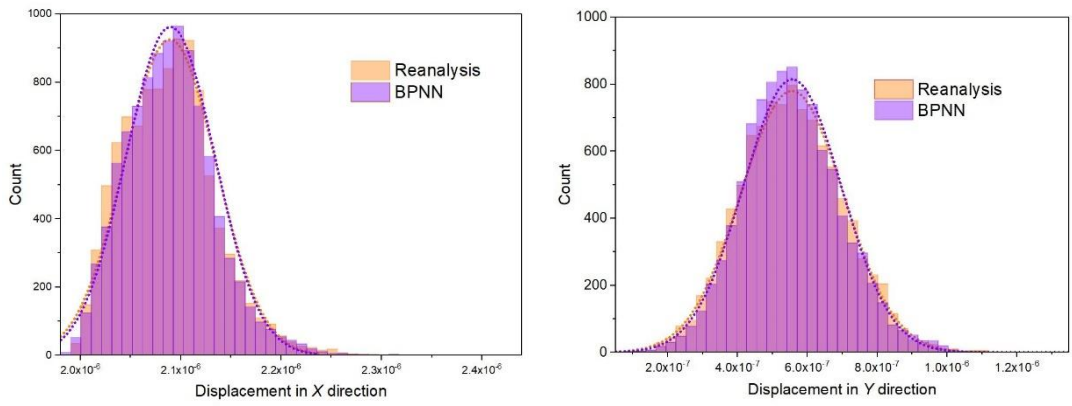

Figure 9. Comparison of PDF of case 1

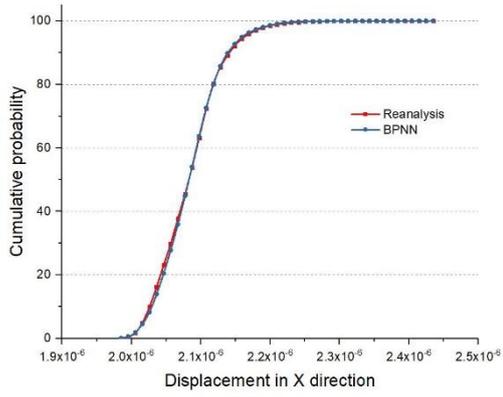 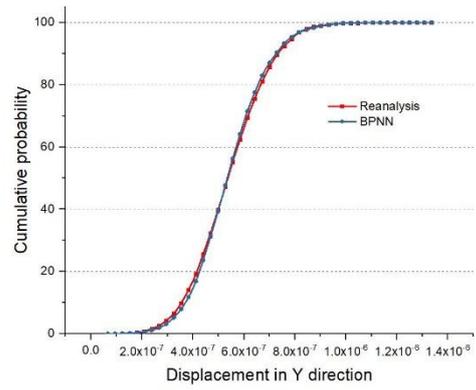

a. CDF comparison in *X* direction  b. CDF comparison in *Y* direction

Figure 10. CDF comparison using normal fit

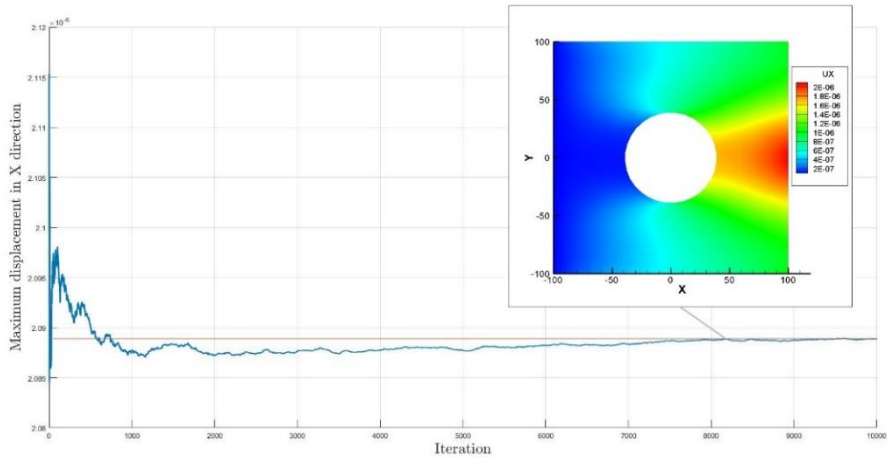

a. MCS convergence and mean response in *X* direction

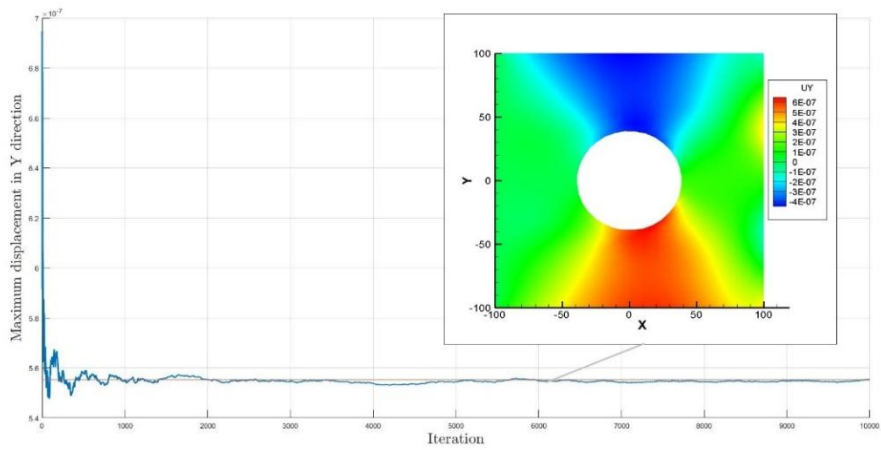

b. MCS convergence and mean response in *Y* direction

Figure 11. Convergence and mean response of case 1

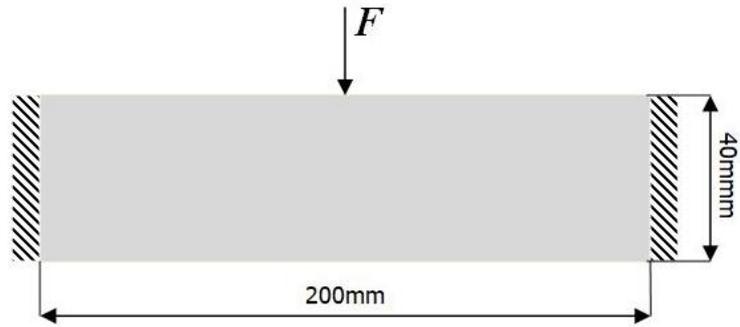

Figure 12. Geometry and boundary condition of beam

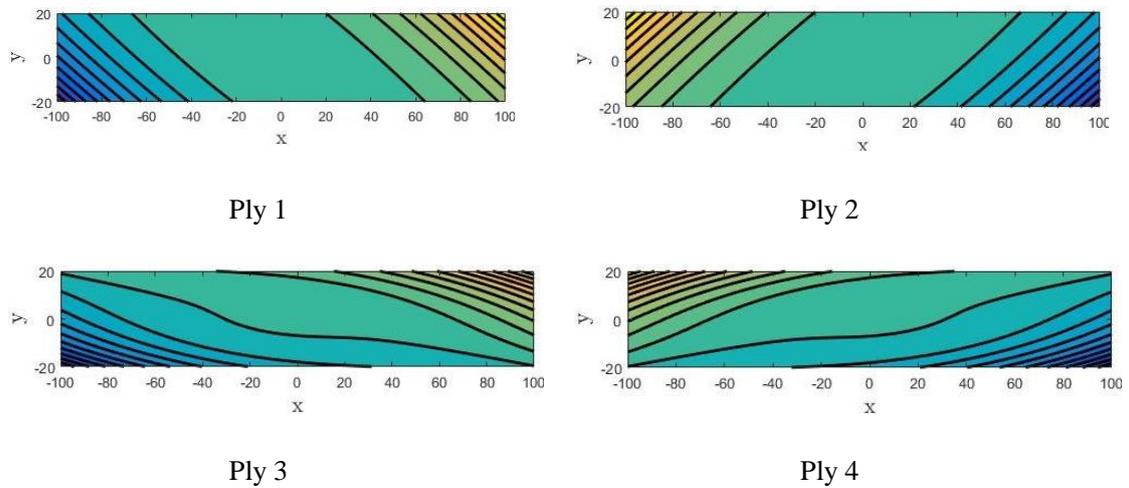

Ply 1          Ply 2

Ply 3          Ply 4

Figure 13. Fiber path of beam

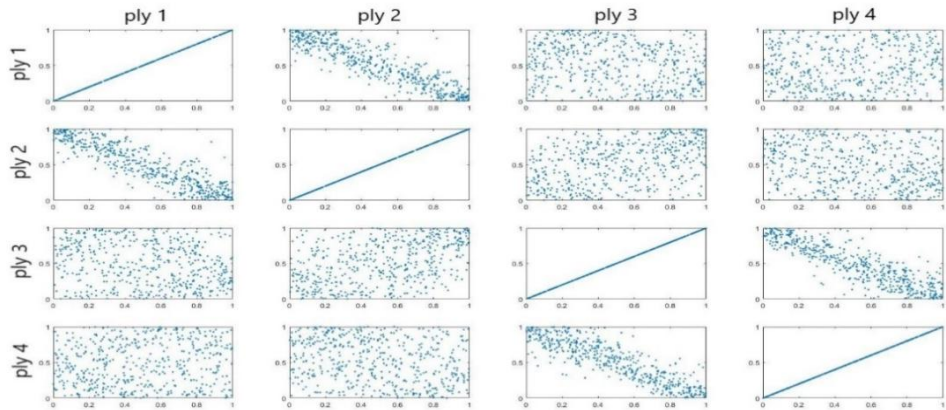

Figure 14. Sample distribution of case 2

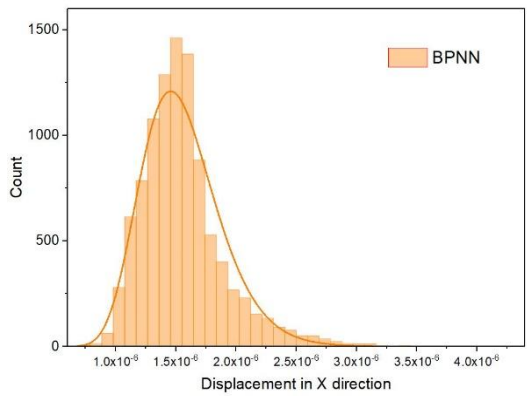

a. Result of BPNN in *X* direction

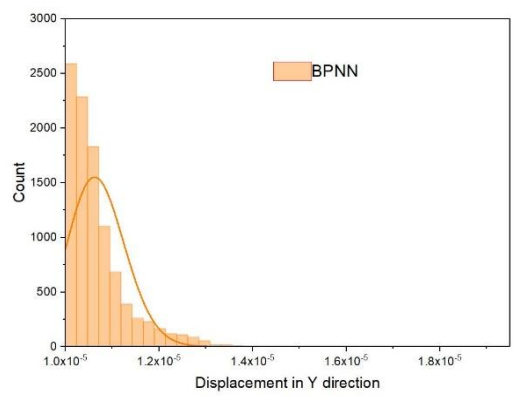

b. Result of BPNN in *Y* direction

Figure 15. PDF of displacement using BPNN of case 2

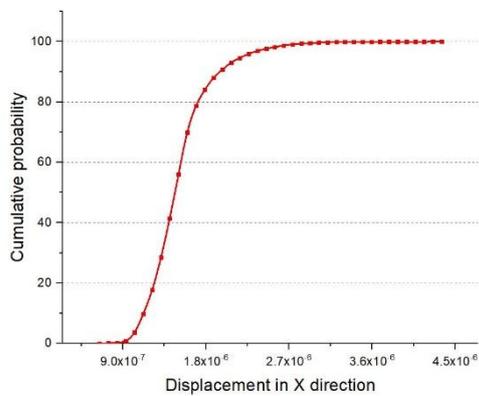

a. CDF in *X* direction

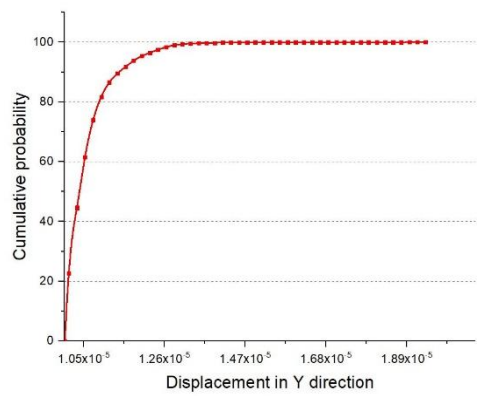

b. CDF in *Y* direction

Figure 16. CDF of displacement using lognormal fit

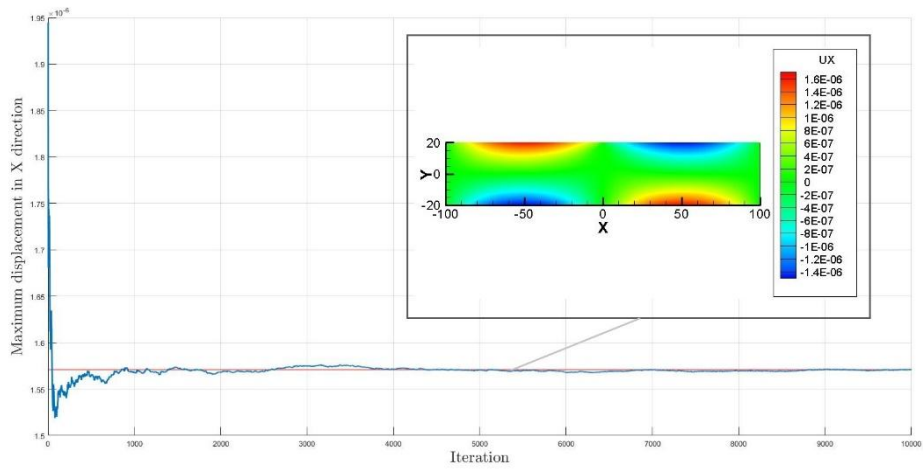

a. MCS convergence and mean response in *X* direction

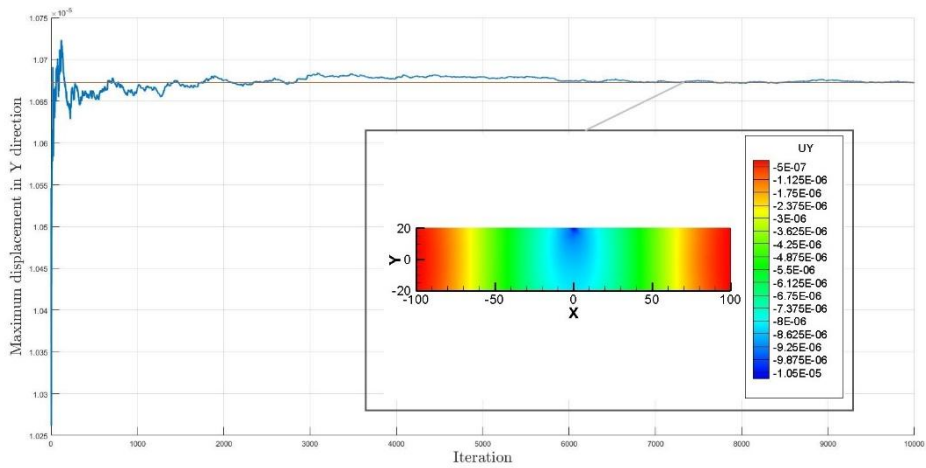

b. MCS convergence and mean response in *Y* direction
Figure 17. Convergence and mean response of case 2

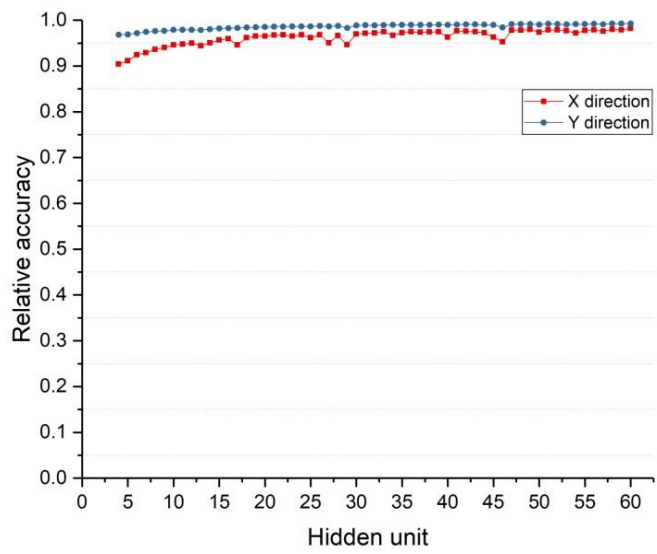

Figure 18. Relation between relative accuracy and hidden units

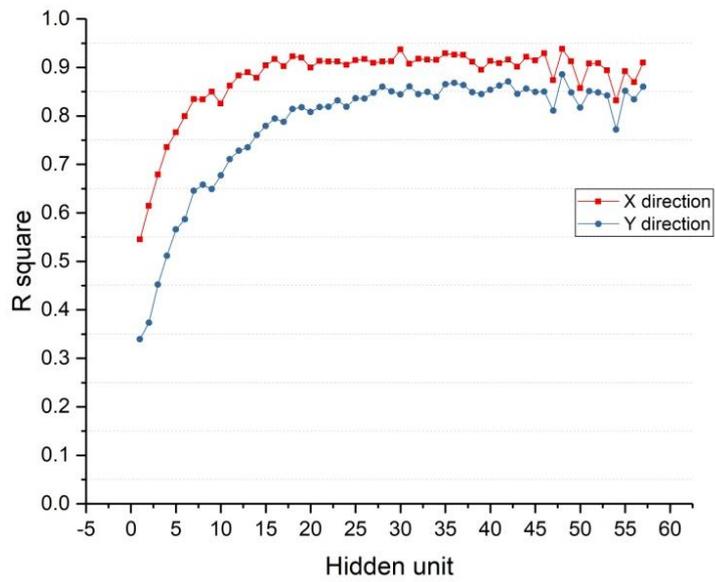

Figure 19. Relation between $R^2$ and hidden unit

Table 1. Material Property

| Material parameters | $E_L$ | $E_T$ | $v_{LT}$ | $G_{LT}$ | $G_{TN}$ | $G_{LN}$ |
|---|---|---|---|---|---|---|
| Value($Kpa$) | 137.9 | 10.34 | 0.29 | 6.89 | 3.9 | 6.89 |

Table 2. Parameters of path function

| Ply | $\alpha_1$ | $\alpha_2$ | $\alpha_3$ | $\alpha_4$ |
|---|---|---|---|---|
| 1 | -2.879 | 0.527 | -0.015 | -9.989 |
| 2 | 2.879 | -0.527 | -0.015 | -9.989 |
| 3 | 6.270 | -0.656 | -1.745 | 7.811 |
| 4 | -6.270 | 0.656 | -1.745 | 7.811 |
| 5 | 7.258 | -0.069 | 4.087 | 17.191 |
| 6 | -7.258 | 0.069 | 4.087 | 17.191 |
| 7 | 14.227 | 2.586 | -2.127 | 10.519 |
| 8 | -14.227 | -2.586 | -2.127 | 10.519 |

Table 3. Mean, variance and bandwidth comparison between two methods of case 1

| Method | X direction | | | Y direction | | |
|---|---|---|---|---|---|---|
| | Mean | Variance | Bandwidth | Mean | Variance | Bandwidth |
| Reanalysis | 2.09E-06 | 2.11E-15 | 4.30E-07 | 5.55E-07 | 2.17E-14 | 1.12E-06 |
| BPNN | 2.09E-06 | 1.98E-15 | 4.08E-07 | 5.56E-07 | 1.99E-14 | 1.24E-06 |
| Relative error | 0.03% | 6.16% | 5.12% | 0.18% | 8.29% | 9.68% |

Table 4. Relative accuracy and $R^2$ of case 1

| Direction | Relative accuracy | $R^2$ |
|---|---|---|
| X | 99.35% | 0.8411 |
| Y | 92.16% | 0.8535 |

Table 5. Parameter of beam path function

| Ply | $a_1$ | $a_2$ | $a_3$ | $a_4$ | $a_5$ | $a_6$ | $a_7$ | $a_8$ |
|---|---|---|---|---|---|---|---|---|
| 1 | 10.611 | -0.5563 | -0.053 | -1.684 | 1.015 | 1.248 | 0.3073 | 0.6024 |
| 2 | 10.611 | 0.5563 | -0.053 | -1.684 | 1.015 | -1.248 | -0.3073 | 0.6024 |
| 3 | 5.402 | 1.846 | 0.3601 | -0.2195 | 0.3203 | 0.9506 | 0.0481 | 2.975 |
| 4 | 5.402 | -1.846 | 0.3601 | -0.2195 | 0.3203 | -0.9506 | -0.0481 | 2.975 |

Table 6. Mean, variance and bandwidth of displacement using BPNN model of case 2

| Direction | Mean | Variance | Bandwidth |
|---|---|---|---|
| $X$ | 1.5567E-6 | 1.097E-13 | 3.5596E-6 |
| $Y$ | 1.0674E-5 | 3.760E-14 | 8.9673E-6 |

Table 7. Efficiency comparison

| Model | Cost of once iteration (s) | | | Totally time saving (s) | |
|---|---|---|---|---|---|
| | Full analysis | Reanalysis | BPNN | Reanalysis | BPNN |
| Hole plate | 8.34 | 1.09 | 3.30E-2 | 7.34E4 | 8.34E5 |
| Plane beam | 1.87 | 2.50E-1 | 2.29E-2 | 1.62E4 | 1.85E4 |